\documentclass[12pt]{article}
\usepackage{graphicx}
\usepackage{epsfig, amsmath, amssymb}

\renewcommand{\bar}[1]{\overline{#1}}

\def\gsim{\mathrel{\rlap{\lower4pt\hbox{\hskip1pt$\sim$}}
\raise1pt\hbox{$>$}}}

\newcommand\la{\langle}
\newcommand\ra{\rangle}
\newcommand\beq{\begin{equation}}

\newcommand\eeq{\end{equation}}
\newcommand\beqn{\begin{eqnarray}}
\newcommand\eeqn{\end{eqnarray}}

\def\GeV{\,\mbox{GeV}}
\def\TeV{\,\mbox{TeV}}
\def\lsim{\mathrel{\rlap{\lower4pt\hbox{\hskip1pt$\sim$}}
\raise1pt\hbox{$<$}}}         
\def\gsim{\mathrel{\rlap{\lower4pt\hbox{\hskip1pt$\sim$}}
\raise1pt\hbox{$>$}}}         

\def\GeV{\,\mbox{GeV}}
\def\MeV{\,\mbox{MeV}}

\begin{document}
\begin{flushright}
USM-TH-214  \\
SLAC-PUB-12664\\
\end{flushright}
\bigskip\bigskip

\centerline{\large \bf Higgs Hadroproduction at Large Feynman \boldmath$x$}

\vspace{22pt}

\centerline{\bf {
Stanley J. Brodsky\footnote{Electronic address:
sjbth@slac.stanford.edu}$^{a}$,
Alfred Scharff Goldhaber\footnote{Electronic address:
goldhab@max2.physics.sunysb.edu}$^{a,b}$,
Boris Z. Kopeliovich\footnote{Electronic address:
bzk@mpi-hd.mpg.de}$^{c,d}$,
Ivan Schmidt\footnote{Electronic address:
ivan.schmidt@usm.cl}$^{c}$}}

{\centerline {$^{a}$SLAC,
{Stanford University, Stanford, CA 94309, USA}}

{\centerline {$^{b}$ C.N. Yang Institute for Theoretical Physics,}
{\centerline {State University of New York, Stony Brook, NY 11794-3840}}

{\centerline {$^{c}$Departamento de F\'{\i}sica y Centro de Estudios
Subat\'omicos,}
{\centerline {Universidad T\'ecnica Federico Santa Mar\'{\i}a,}
{\centerline {Casilla 110-V, Valpara\'\i so, Chile}}

{\centerline {$^{d}$Joint Institute for Nuclear Research, Dubna, Russia}}

\vspace{10pt}
\begin{center}
{\large \bf Abstract}
\end{center}
We propose a novel mechanism for the production of the Higgs boson
in inclusive hadronic collisions, which utilizes the presence of
heavy quarks in the proton wave function. In these inclusive
reactions the Higgs boson acquires the momenta of both the heavy
quark and antiquark and thus carries 80\% or more of the
projectile's momentum. We predict that the cross section ${d
\sigma/d x_F }(p \bar p \to H X)$ for the inclusive production of
the Standard Model Higgs coming from intrinsic bottom Fock states is
of order 150 fb  at LHC energies, peaking in the region of $x_F \sim
0.9$. Our estimates indicate that the corresponding cross section
coming from gluon-gluon fusion at $x_F = 0.9$  is relatively
negligible and therefore the peak from intrinsic bottom should be
clearly visible for experiments with forward detection capabilities.
The predicted cross section for the production of the Standard Model
Higgs coming from intrinsic heavy quark Fock states in the proton is
sufficiently large that detection at the Tevatron and the LHC may be
possible.
\newpage

\section{Introduction}

Theoretical predictions for the production of the Higgs at the
Tevatron and the LHC, and the relevant QCD backgrounds, have been
extensively developed~\cite{Buttar:2006zd,DelDuca:2006wb}. The main
hadroproduction mechanisms for $ p p (\bar p) \to H X$ are gluon
fusion $gg \to H$ through the top quark triangle loop, weak-boson
fusion subprocesses such as $q \bar q \to q \bar q W W \to q \bar q
H,$  Higgs-strahlung processes $q \bar q \to W(Z) H,$ and associated
top pair production $g g \to t \bar t H,$ which again utilizes the
large coupling of the Higgs to the heavy top quark.   A common
characteristic of these reactions is the strong dominance of the
production cross section ${d\sigma/ d x^H_F}$ at central rapidities,
i.e. $x^H_F = {2 p^H_L/\sqrt s} \simeq 0,$ particularly for the
reactions initiated by the gluon distribution in the colliding
proton or antiproton.  In each case the cross section falls as a
power of $(1-x^H_F)^n$ at large $x^H_F$,  $n \simeq 3 \to 5$, as one
approaches the fragmentation regions.

In this paper we will discuss a novel QCD mechanism for Higgs
hadroproduction in which the Higgs is produced at large $x^H_F \ge
0.8$, a regime where we expect that backgrounds from other QCD and
Standard Model processes should be small.

One can demonstrate from the operator product
expansion~\cite{Franz:2000ee} that the proton has finite probability
for its wavefunction to contain intrinsic heavy flavors $s \bar s, c
\bar c, b \bar b, t \bar t$ through its quantum fluctuations. The
dominant $|uud Q \bar Q>$ bound-state configuration occurs when the
wavefunction is minimally off its energy-shell; i.e., the most
likely configuration occurs when all of the constituents have  the
same rapidity~\cite{Brodsky:1980pb,Brodsky:1981se}. In terms of the
light-front fractions $x_i = {k^+_i/ P^+}$, the light-front
wavefunction of a hadron is maximal when the constituents have
light-front momentum fractions $x_i  \propto \sqrt{k^2_{\perp i}+
m^2_i}$ with $\sum_i x_i =1.$ Thus  the heavy $Q$ and $\bar Q$
quarks in the $|uudQ\bar{Q}>$ Fock state will carry most of the
proton's momentum.  A typical configuration is $x_Q \sim x_{\bar Q }
\sim 0.4$ and $x_q \sim 0.07$.

A test of intrinsic charm is the measurement of the charm quark
distribution $c(x,Q^2)$ in deep inelastic lepton-proton scattering
$\ell p \to \ell^\prime c X.$  The data from the European Muon
Collaboration (EMC) experiment~\cite{Aubert:1982tt}, show an excess
of events in the charm quark distribution at the largest measured
$x_{bj}$,
 beyond predictions
based on gluon splitting and DGLAP evolution. Next-to-leading order
(NLO) analyses~\cite{Harris:1995jx} show that an intrinsic charm
component, with probability of order $1\%$, is allowed by the EMC
data in the large $x_{bj}$ region. This value is consistent with an
evaluation based on the operator product
expansion~\cite{Franz:2000ee}. Although these estimates still have
large uncertainties~\cite{Watt:2008hi}, our calculations in what
follows will be based on this number as a best scenario.

The importance of further direct
measurements of the charm and bottom distributions at high $x_{bj}$
in deep inelastic scattering has  been stressed by Pumplin, Lai, and
Tung~\cite{Pumplin:2007wg,Pumplin:2005yf}. These authors also give a
survey of intrinsic charm models derived from perturbation theory
and non-perturbative theory (based on meson-nucleon fluctuations),
as well as the current experimental constraints from deep inelastic
lepton scattering.


As noted in ref.~\cite{Brodsky:2006wb}, the presence of high-$x$
intrinsic heavy quark components in the proton's structure function
will also lead to Higgs production at high $x_F$ through
subprocesses such as $g b \to H b$; such reactions could be
particularly important in MSSM models in which the Higgs has
enhanced couplings to the $b$ quark~\cite{Nilles:1983ge}.

In this paper we shall show how one can utilize the combined high
$x$ momenta of the $Q $ and $ \bar Q$ pair  to produce the Higgs
inclusively in the reaction $p p \to H X$ with 80\% or more of the
beam momentum.  The underlying subprocess which we shall utilize is
$ (Q \bar Q) g  \to H. $

The probability for intrinsic heavy quark Fock states is suppressed
as $\Lambda^2\over m^2_Q$, corresponding to the power-falloff of the
matrix element of the effective non-Abelian twist-six  $g g \to g g$
operator $<p|G^3_{\mu \nu} |p>$ in the proton
self-energy~\cite{Franz:2000ee}.  This behavior also has been
obtained in explicit perturbative calculations~\cite{Boris}.
However, the $1\over m^2_Q$ suppression in the intrinsic heavy quark
probability is effectively compensated by the Higgs coupling to the
heavy quark, which is of order $G_F m^2_Q$ in the rate. In fact, as
we shall show, production from the heavy intrinsic $|uud b \bar b >$
Fock state is also enhanced due to the increased probability of its
overlap with the small-size Higgs wavefunction in comparison with
$|uud c \bar c>$. Since $x^H_F = x_1(Q \bar Q) - x_g$,  the Higgs is
produced with approximately the sum of the $Q$ and $\bar Q$ momenta,
and therefore the Higgs hadroproduction cross section ${d
\sigma\over d x_F}( p p \to H X)$ benefits from intrinsic charm,
bottom, and even intrinsic top fluctuations in the projectile
hadron.

It is also important to note that the dominant $|uudQ\bar{Q}>$ Fock
state of the projectile proton, arising from the  cut of the $G^3$
non-Abelian operator in the proton self energy, requires  the $Q
\bar Q$  pair to be in a color-octet configuration.  Thus the $Q
\bar Q$ pair in the proton $|(uud)_{8C}  (Q\bar Q)_{8C}>$ Fock state
can be  converted,  by a single gluon exchange  with the target in
the hadronic collision, to a color-singlet  Higgs  through the
matrix element $<(Q \bar Q)_{8C} |G^{\mu \nu}|H>$.   A second gluon
exchange with the target nucleon allows one to produce the Higgs
diffractively or semi-inclusively: $ p p \to H + p + X$, where the
final state proton from the target is produced isolated in rapidity~\cite{Brodsky:2006wb}.

Intrinsic heavy quark Fock states can thus be remarkably efficient
in converting collision energy to Higgs production energy. We thus
advocate searching for the Higgs at values of $x_F$ as large as or
beyond 0.8. The characteristic shape of the IQ Higgs signal will be
a peak at large $x_F$, sitting on the rapidly falling power-law
suppressed contributions from  central rapidity Higgs
hadroproduction processes such as $ g g \to H.$  QCD backgrounds to
the $H \to b \bar b$ decay from subprocesses such as $ g g \to b
\bar b$ will also be power law suppressed at high $x_F.$

It is interesting that in principle it is possible to use the $7 $
TeV LHC beam and the $g Q \bar Q \to H$ subprocess in a fixed target
mode $ p A \to H X$, which together with nuclear Fermi momentum, can
lead to light Higgs production at threshold.

In a previous paper~\cite{Brodsky:2006wb} we considered doubly
diffractive Higgs production $ p p \to p + H +p$, arising from the
heavy quark Fock state component, where the Higgs is produced in the
fragmentation region of one of the colliding protons. The presence
of a large rapidity gap and the possibility to perform a missing
mass measurement which only needs detection of the  two final-state
protons substantially increases the chances to detect the high $x_F$
Higgs particle, since its decay to a specific channel is not
required. In this case the magnitude of the cross section of Higgs
production reaches a level which can be measured at LHC with a
proper trigger system.

In this paper we will estimate the cross section for inclusive Higgs
production, both at the LHC and the Tevatron, arising from the
proton heavy quark Fock state. Although we do not have the advantage
of the missing mass experiment as in the exclusive
doubly-diffractive case, the cross section itself is considerably
larger.  Since the Higgs can acquire  most of the momentum of the
intrinsic heavy quark pair, it will be produced at quite large $x_F$
values, where the background from multi-particle production is
expected to be small. Nevertheless, the main goal of this paper is
to evaluate the cross section of this novel Higgs production
mechanism, leaving a detailed clarification of the relevant
backgrounds to further investigations.

\section{ Inclusive Hadroproduction of the Higgs from Intrinsic Heavy Flavors}

We start our analysis with the totally inclusive case, described by the diagram
of Fig.~\ref{QQ-inclusive}.
\begin{figure}[tbh]
\includegraphics{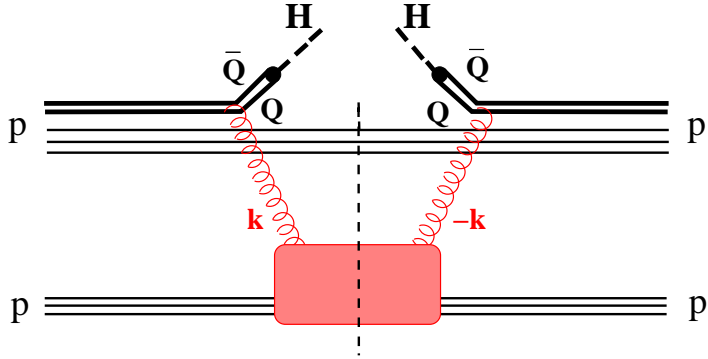}
\begin{center}
\vspace{6.5cm}
\parbox{13cm}
{\caption[Delta] {Representation of the cross section for inclusive Higgs production.
The dashed line shows the unitarity cut.} \label{QQ-inclusive}}
\end{center}
\end{figure}
Since the remnants of both protons are color octets, no rapidity gap is
expected.

Assuming that the intrinsic $\bar QQ$ pair is in a color-octet
$P$-wave state, the corresponding cross section can be readily
obtained.  It is given by
\beqn \frac{d\sigma}{d^2k\,dz} &=&
P_{IQ}(z)\, \frac{2\pi\alpha_s(k^2)}{3}\, \frac{{\cal
F}(x,k^2)}{k^4} \left| \int d^2r H^\dagger(\vec r) e^{-i\vec
k\cdot\vec r/2} \left(1-e^{i\vec k\cdot\vec r}\right)\,
\Psi_{\bar QQ}(\vec r)\right|^2\nonumber\\
&=& P_{IQ}(z)\,G_F\,
\frac{\alpha_s(k^2)
{\cal F}(x,k^2)}{k^2\,\delta^2}
\times
\left\{\begin{array}{cc}
\ln(\delta)&{\rm if}\ Q=c,\ b\\
{1\over 8}\left[1+
\frac{1-\delta}{\delta}\,\ln(1-\delta)\right]&
{\rm if}\ Q=t
\end{array}
\right. \label{10} \eeqn
Here $\vec k$ and $z=x_F$ are the
transverse momentum and longitudinal momentum fraction of the Higgs
particle, and $\delta=M_H^2/4m_Q^2$. We assume that $M_H<2m_t$.  In
this expression $P_{IQ}(z)$ is the heavy quark pair distribution
function which will be specified later, $H(\vec r)$ is the relevant
light-front wave function of the Higgs in the impact parameter
representation, $G_F$ is the Fermi constant, $\Psi_{\bar QQ}$ is the
$\bar QQ$ wave function of the $Q \bar Q$ pair in the proton, and
${\cal F}(x,k^2)= \partial G(x,k^2)/\partial({\rm ln}k^2)$ is the
unintegrated gluon density, where $G(x,Q^2)=x\,g(x,Q^2)$, and
$x=M_H^2/zs$. The specific form of these functions will be discussed
below.

Notice that the form of Eq.~(\ref{10}) (upper line) is closely
related to the well known dipole-proton inelastic cross section,
which helps to fix the structure of (\ref{10}) and the pre-factors.
Indeed, if to integrate in (\ref{10}) over $z$ and  $k^2$, and
instead of projecting to the Higgs wave function to sum over all
final states $X(\vec r)$ of the $\bar QQ$, due to completeness
$\sum\limits_X X(r)X^*(\vec r^{\,\prime})=\delta(\vec r-\vec
r^{\,\prime})$. Then one arrives at the time reversal $\bar
QQ$-proton inelastic cross section, \beq \sigma^{\bar
QQ-p}_{in}=\int d^2r\,|\Psi_{\bar QQ}(r)|^2\,\sigma(r,x), \label{12}
\eeq where the dipole-proton cross section has the saturated form
(see e.g. in \cite{gbw}),
 \beq
 \sigma(r,x)=\frac{4\pi}{3}\int \frac{d^2k}{k^2}\,
 \alpha_s(k^2)\,{\cal F}(x,k^2)
 \left(1-e^{i\vec k\cdot\vec r}\right).
 \label{14}
 \eeq

\section{Elements of the Calculation}

\subsection{The Unintegrated Gluon Density}

The unintegrated gluon density, ${\cal F}(x,k^2)=
\partial G(x,k^2)/\partial({\rm ln}k^2)$, is the
transverse momentum differential of $G(x,Q^2)=x\,g(x,Q^2)$. It
preserves the infrared stability of the cross section, since ${\cal
F}$ vanishes at $k^2\to 0$.  The phenomenological gluon density
fitted to data includes by default all higher order corrections and
supplies the cross section with an energy dependence important for
the extrapolation to very high energies. One can relate the
unintegrated gluon distribution to the phenomenological dipole cross
section fitted to data for $F_2(x,Q^2)$ from HERA, as was done in
Ref.~\cite{gbw}, \beq {\cal F}(x,k^2) =
\frac{3\,\sigma_0}{16\,\pi^2\,\alpha_s(k^2)}\ k^4\,R_0^2(x)\, {\rm
exp}\Bigl[-{1\over4}\,R_0^2(x)\,k^2\Bigr]\ . \label{1410} \eeq Here
$R_0(x)=0.4(x/x_0)^{0.144}$ fm (with $x_0 = 3 \times 10^{-4}$), and
$\sigma_0 = 23.03$ mb~\cite{gbw}.

\subsection{The Higgs Wavefunction}

The $P$-wave LF wave function of the Higgs in impact parameter
representation is given by the Fourier transform of its Breit-Wigner
propagator: \beq H(\vec r) = i\,\frac{\sqrt{3\, G_F}}{2\pi}\, m_Q\,
\bar\chi\,\vec\sigma\,\chi\, \frac{\vec r}{r}\,
\left[\epsilon\,Y_1(\epsilon r)-{ir\over2}\,\Gamma_HM_H\,
Y_0(\epsilon r)\right] \ . \label{1500} \eeq This is the effective
LF form factor representing the effective $L=1$  coupling of the
scalar Higgs to $c \bar c$ or $b \bar b$. Here $G_F$ is the Fermi
constant, $\chi$ and $\bar\chi$ are the spinors for $c$ and $\bar c$
respectively and \beq \epsilon^2=\alpha(1-\alpha)M_H^2 - m_Q^2\ ,
\label{1505} \eeq where $\alpha$ is the fraction of the LF momentum
of the Higgs carried by the $c$-quark. The functions $Y_{0,1}(x)$ in
Eq.~(\ref{1500}) are the second-order Bessel functions and
$\Gamma_H$ is the total width of the Higgs.  If we assume
$\Gamma_H\ll M_H$, we can neglect the second term in
Eq.~(\ref{1500}).

The LF wavefunction Eq.~(\ref{1500}) assumes that the Higgs mass is
much larger than the quark masses, which is  true for the charm
and bottom quarks. However, since it is expected that $2m_t>M_H$,  the
Higgs wave function in this case has the form \beq H_{\bar tt}(\vec r) =
\frac{\sqrt{3\, G_F}}{2\pi}\, m_t\, \bar\chi\,\vec\sigma\,\chi\,
\frac{\vec r}{r}\,\epsilon_t\,K_1(\epsilon_t r)\ , \label{1506} \eeq
where $K_1(x)$ is the modified Bessel function and \beq
\epsilon_t^2=m_t^2-\alpha(1-\alpha)M_H^2~~. \label{1506a} \eeq

The probabilities computed from the wave functions Eqs.~(\ref{1500})
and (\ref{1506a}) require regularization in the ultraviolet limit
\cite{krt,e-loss}, as is the case of the $\bar qq$ wave function of
a transverse photon. Such wave functions are not solutions of the
Schr\"odinger equation, but are distribution functions for
perturbative fluctuations. They are overwhelmed by very heavy
fluctuations with large intrinsic transverse momenta or vanishing
transverse separations. Such point-like fluctuations lead to a
divergent normalization, but they do not interact with external
color fields, i.e., they are not observable. All of the expressions for
measurable quantities, including the cross section, are finite.

\subsection{Modeling the Heavy-Quark Intrinsic Distribution}

The intrinsic heavy quark contribution arises in perturbation theory
from the unitary cut of the $ g g \to Q \bar Q \to g g $ box diagram
in the proton self-energy, the analog of light-by light scattering
in QED. The gluons are attached to the valence quarks of the proton.
The QCD box graph can be computed perturbatively and it is the basis
for the perturbative analysis of intrinsic heavy quark Fock states.
In non-Abelian theory this contribution falls as $\Lambda^2/m^2_Q$.
The same result is obtained from the operator product expansion
analysis of Franz, Polyakov, and Goeke~\cite{Franz:2000ee}. The
perturbative IQ analysis has a dual representation in terms of heavy
meson-baryon Fock states. In the case of the intrinsic charm and
bottom Fock states $| u u d Q \bar Q>$, it is reasonable to assume
that the quarks cluster, so that the IQ wavefunction resembles a
meson-baryon fluctuation such as $|\Lambda_c D>$ or $|\Lambda_b B>$.
Similar duality arguments can be used to model the $|u u d s \bar
s>$ state~\cite{Burkardt:1991di,Brodsky:1996hc}. This is the basis
of our ``nonperturbative" model of IQ. This duality argument is not
applicable to the heavy, short-lived $| u u d t \bar t>$ state, so
we have relied on the perturbative IQ model in this case. The total
probability for the IQ Fock state in either case varies as
$\Lambda^2/m^2_Q$, as guaranteed by the OPE.

In contrast to the IQ mechanism, the perturbative gluon-splitting $
g \to Q \bar Q $ contribution to the heavy quark structure function
of the proton depends logarithmically on the quark mass in a hard
reaction. A gluon can thus fluctuate into $Q \bar Q$  with
essentially no suppression. However to be produced on mass shell the
fluctuation must interact, and the interaction cross section is
$1/m_Q^2$ suppressed due to color transparency as in heavy quark
production in deep inelastic lepton scattering. This is explicitly
shown by Floter et al.~\cite{Floter:2007xv}. This mechanism
underlies the usual gluon fusion contribution to Higgs
hadroproduction at small $x_F$ through the heavy quark triangle
graph.

We shall thus assume the presence in the proton of an intrinsic
heavy quark (IQ) component, a $\bar QQ$ pair, which is predominantly
in a color-octet state.  We will consider two models, corresponding
to the nonperturbative or perturbative origin of the intrinsic heavy
quark Fock state $|uud Q \bar Q>$.

In the nonperturbative model, the heavy component $Q \bar Q$ will
interact strongly with the remnant $3q$ valence quarks. Such
nonperturbative reinteractions of the intrinsic sea quarks in the
proton wavefunction can lead to a $Q(x) \ne \bar Q(x)$ asymmetry as
in the $\Lambda K$ model for the $s \bar s$
asymmetry~\cite{Burkardt:1991di,Brodsky:1996hc}. As is the case for
charmonium, the mean $\bar QQ$ separation is expected to be
considerably larger than the transverse size $1/m_Q$ of perturbative
$\bar QQ$ fluctuations. We thus will assume that the nonperturbative
wave function for the $Q \bar Q$ is  an S-wave solution of the
Schr\"odinger equation. Assuming an oscillator potential we have
\beq \Psi^{npt}_{\bar QQ}(\vec r) = \sqrt{\frac{m_Q\omega}
{2\pi}}\,\exp(-r^2\,m_Q\,\omega/4)\ , \label{1510} \eeq where
$\omega$ is the oscillation frequency. The mean heavy inter-quark
distance is \beq \la r_{\bar QQ}^2\ra = \frac{2}{\omega\,m_Q}\ .
\label{5} \eeq
 We will give further details on the nonperturbative
model in the next subsection.

Alternatively, the  IQ component can be considered to derive
strictly perturbatively from the minimal gluonic couplings of the
heavy-quark pair to two valence quarks of the proton; this is likely
the dominant mechanism at the largest values of
$x_Q$~\cite{Brodsky:1991dj}.  In this case the transverse separation
of the $\bar QQ$ is controlled by the energy denominator, $\la
r_{\bar QQ}^2\ra = 1/m_Q^2$, which is much smaller in size than the
estimate given by Eq.~(\ref{5}) for the nonperturbative model.

Since the Higgs is produced from an $S$-wave $\bar QQ$, the
perturbative distribution amplitude is ultraviolet stable.  We can
normalize  $P_{IQ}$ to the probability to have such a heavy quark pair
in the proton and take the perturbative wavefunction as
\beq \Psi^{pt}_{\bar QQ}(\vec r) = \frac{m_Q}
{\sqrt{\pi}}\,K_0(m_Q r)\ . \label{1520} \eeq
Here the modified
Bessel function $K_0(m_Q r)$ is the Fourier transform of the energy
denominator associated with the $\bar QQ$ fluctuation, given by equation (\ref{1800}) below. We assume the
$Q$ and $\bar Q$ quarks carry equal fractional momenta. For fixed
$\alpha_s$,  the energy denominator governs the probability of the
fluctuation in momentum space, since the charm quarks are treated as
free particles in the perturbative analysis.

\subsubsection{Nonperturbative Intrinsic Heavy Quarks--Further Details}

In principle, one can construct the complete Fock state
representation of the hadronic LF wave function by diagonalizing the
LF Hamiltonian. Here we will model the intrinsic heavy quark
component by using the method of Ref.~\cite{terentiev} in which one
uses a Lorentz boost of a wave function assumed to be known in the
hadron rest frame. The Lorentz boost will also generate higher
particle number quantum fluctuations which are missed by this
procedure; however, this method is known to work  well in some
cases~ \cite{hikt,kth}, and even provides a nice cancelation of
large terms violating the Landau-Yang theorem~ \cite{kt}.

We shall assume that the rest frame intrinsic heavy quark  wave function has the Gaussian
form (in momentum space),
\beq
\tilde\Psi_{IQ}(\vec Q,z) =
\left(\frac{1}{\pi\omega\mu}\right)^{3/4} \exp\left(-\frac{\vec
Q^2}{2\omega\mu}\right)\ .
\label{2000}
\eeq
Here $\omega\simeq 300 \MeV$ stands for the oscillator frequency and
$\mu=M_{\bar QQ}M_{3q}/(M_{\bar QQ}+M_{3q})$ is the reduced mass
of the $\bar QQ$ and $3q$ clusters. For further estimates we use
$M_{\bar QQ}=3\GeV$ and $M_{3q}=1\GeV$, although the latter could
be heavier, since it is the $P$-wave.

We can relate the 3-vector $\vec Q$ to the effective mass of the
system, $M_{eff}=\sqrt{\vec Q^2 +M_{\bar QQ}^2}+ \sqrt{\vec Q^2
+M_{3q}^2}$, by using LF variables, $\vec Q$ and $z$, \beq
M_{eff}^2=\frac{Q_T^2}{z(1-z)}+\frac{M_{\bar QQ}^2}{z}+
\frac{M_{3q}^2}{1-z}\ , \label{2020} \eeq where $Q_T$ is Q´s
transverse component. Then the longitudinal component $Q_L$ in the
exponent in (\ref{2000}) reads, \beq Q_L^2=\frac{M_{eff}^2}{4}+
\frac{(M_{\bar QQ}^2-M_{3q}^2)^2} {4M_{eff}^2} - \frac{M_{\bar
QQ}^2+M_{3q}^2}{2} - Q_T^2\ , \label{2040} \eeq and the LF wave
function acquires the form, \beq \Psi_{IQ}(Q,z) = K\,
\exp\left\{-\frac{1}{8\omega\mu}\left[M_{eff}^2 + \frac{(M_{\bar
QQ}^2-M_{3q}^2)^2}{M_{eff}^2} \right]\right\}\ , \label{2060} \eeq
where \beqn K^2 &=& \frac{1}{8Q_L}\,
\left(\frac{1}{\pi\omega\mu}\right)^{3/2}\, \exp\left(\frac{M_{\bar
QQ}^2+M_{3q}^2}{2\omega\mu}\right)\, \left[1-\frac{(M_{\bar
QQ}^2-M_{3q}^2)^2} {M_{eff}^4}\right] \nonumber\\ &\times&
\left[\frac{Q_T^2(2z-1)}{z^2(1-z)^2} - \frac{M_{\bar QQ}^2}{z^2}+
\frac{M_{3q}^2}{(1-z)^2}\right]~. \label{2080} \eeqn
We can now calculate the $z$-dependence of the function
$P_{IQ}(z)$ introduced in Eq.~(\ref{10})),
\beq
\frac{P_{IQ}(z)}{P_{IQ}}= \
\int d^2Q\,\left|\Psi_{IQ}(Q,z)\right|^2~. \label{2100} \eeq We
shall assume $P_{IQ}=0.01$ for intrinsic charm, and a suppression
factor proportional to $1/m^2_Q$ for heavier flavors.

We have included the corresponding evolution of this
non-perturbative distribution from the bound state scale up to the
Higgs scale, whose effect is to bring in a factor of $(1-z)^{\Delta
p}$, to be discussed in detail later in the paper.

\subsubsection{Perturbative Intrinsic Heavy Quarks--Further Details}

The light-front wave function of a perturbative fluctuation $p\to
|3q\bar QQ\ra$ in momentum representation is controlled by the
energy denominator, \beq \Psi_{IQ}(Q,z,\kappa)\propto
\frac{z(1-z)}{Q^2+z^2m_p^2+M^2_{\bar QQ}(1-z)}\ . \label{1800} \eeq
Here $\vec Q$ is the relative transverse momentum of the $3q$ and
$\bar Q Q$ clusters of the projectile. The effective mass of the
$\bar Q Q$ pair depends on its transverse momentum ($\vec{\kappa}$)
 $M_{\bar QQ}^2=4(\kappa^2+m_Q^2)$.
 %
%
As we have discussed in the introduction, one can show from first
principles using the operator product expansion~\cite{Franz:2000ee}
that the probability of the intrinsic heavy quark Fock state $|uud
\bar Q Q>$ in the proton wavefunction scales as $1\over m^2_Q$. This
power behavior results from integration over the square of the
light-front wavefunction over the invariant mass of the $uud \bar Q
Q$ quarks, and thus dictates its power-law fall-off.

 The Higgs production amplitude is
controlled by the convolution of the IQ $\bar Q Q$ wave function
with the $P$-wave $\bar QQ$ wave function of the Higgs together with
the one-gluon exchange amplitude.  The result has the form
 \beqn && \int\limits_{0}^{\infty} d\kappa^2\,
\Psi_{IQ}(Q,z,\kappa)\left[H_{\bar QQ}(\vec\kappa+\vec k/2)- H_{\bar
QQ}(\vec\kappa-\vec k/2)\right] \nonumber \\ &\propto& z(1-z)\,
\frac{\ln\left[\frac{\left|M_H^2-4m_Q^2\right|(1-z)}
{Q^2+4m_Q^2(1-z)+m_p^2z^2}\right]} {M_H^2(1-z)+Q^2+m_p^2z^2}\ .
\label{1820} \eeqn
This expression peaks at $1-z\sim m_p/M_H$;
therefore the logarithmic factor hardly varies as a function of
$Q^2$. (In the semi-inclusive case, the momentum transfer to the
target proton  is restricted by its form factor.) Making use of this
we perform the integration in Eq.~(\ref{2100}) and arrive at the
following $z$-distribution,
\beq \frac{P_{IQ}(z)}{P_{IQ}}\propto z(1-z)\,
\frac{\left\{\ln\left[\frac{\left|M_H^2-4m_Q^2\right|(1-z)}
{4m_Q^2(1-z)+m_p^2z^2}\right]\right\}^2} {M_H^2(1-z)+m_p^2z^2}\ ,
 \eeq
 This result is actually valid at the scale of
about $4m_Q^2$, and it should be evolved to the scale of the Higgs
mass squared. In order to do that we multiply the distribution by a
factor $(1-z)^{\Delta p}$, where $\Delta p = p(M_H^2) - p(4m_Q^2)$,
and
\beq p(Q^2) = (4C_A/\beta_1) ln[ln(Q^2/\Lambda^2)], \eeq
where
$C_A = 3$, because the intrinsic heavy quark pair is in a color octet state,
and $\beta_1 = 11 - 2/3 n_f$ \cite{BBS}. This is a kind
of Sudakov factor suppression \cite{KPS}. Since the gluon splitting
functions are universal, this is a correct procedure. So finally we
get:
\beq \frac{P_{IQ}(z)}{P_{IQ}} = N z(1-z)^{(1+\Delta p)}\,
\frac{\left\{\ln\left[\frac{\left|M_H^2-4m_Q^2\right|(1-z)}
{4m_Q^2(1-z)+m_p^2z^2}\right]\right\}^2} {M_H^2(1-z)+m_p^2z^2}\ ,
\label{1830} \eeq where $N$ is a constant normalizing  the integral
over $z$ to unity. For the case of the charm quark we obtain $\Delta
p = 1.41$, while for the bottom quark we get $\Delta p = 0.88$.

\section{Results}

\begin{figure}[tbh]
\includegraphics{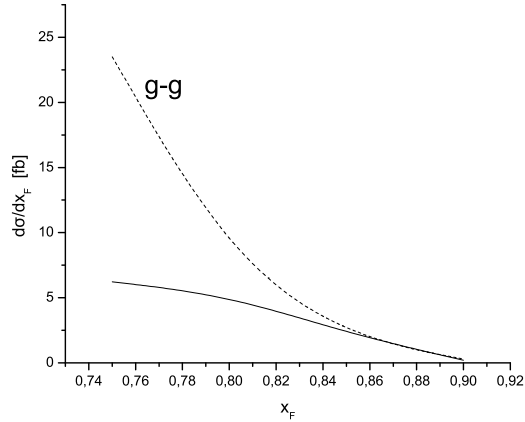}
\begin{center}
\vspace{6.5cm}
\parbox{13cm}
{\caption[Delta] {The cross section of inclusive Higgs production in
$fb$, coming from the non-perturbative intrinsic charm distribution,
at LHC ($\sqrt{s}=14\TeV$) energies.  For comparison we show
also an estimate of the cross section for gluon-gluon fusion.} \label{Inc-NP-IC-LHC}}
\end{center}
\end{figure}

We present in Fig.~\ref{Inc-NP-IC-LHC} the prediction for the
inclusive Higgs production cross section $d \sigma/dX_F( p \bar p
\to H X)$ coming from the intrinsic non-perturbative charm
distributions at LHC ($\sqrt{s}=14\TeV$) energies. It turns out that
the magnitude of this cross section is considerably smaller than the
corresponding gluon-gluon fusion cross section, so we turn to
intrinsic bottom.

We can compare our predictions for inclusive Higgs production coming
from IC with an estimate for the Higgs production gluon-gluon fusion
process. For this we use the following procedure. Using the NNLO
results of ref. \cite{Anastasiou:2004} we see that $d\sigma_H/dy = 1
pb$ at $y=3$, which corresponds to $x_F=0.165$. Then, since

$${d\sigma_H \over dx_F}={e^{-y} \sqrt{s} \over M_H}{d\sigma_H \over dy}$$
we get
$${d\sigma_H \over dx_F}(x_F=0.165)= 5.83 pb.$$
We extrapolate to higher values of $x_F$ using the MRST2006NNLO
gluon distributions at the Higgs scale \cite{Martin:2007bv}, and
calculate the suppression factors due to rise of $x_1=x_F$. At the
same time $x_2$ decreases and the gluon density in the target rises
($x_2=m_H^2/s/x_1$). Eventually we get
gg-cross section shown in Fig. \ref{Inc-NP-IC-LHC}.


\begin{figure}[tbh]
\includegraphics{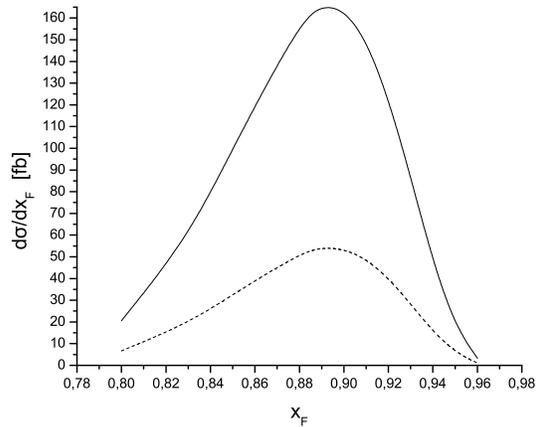}
\begin{center}
\vspace{6.5cm}
\parbox{13cm}
{\caption[Delta] {The cross section of inclusive Higgs production in
$fb$, coming from the nonperturbative intrinsic bottom distribution,
at both LHC ($\sqrt{s}=14\TeV$, solid curve) and Tevatron
($\sqrt{s}=2\TeV$, dashed curve) energies.}
\label{Inc-NP-IB-LHC}}
\end{center}
\end{figure}

The same analysis can be repeated for inclusive Higgs production
from intrinsic bottom (IB), assuming that the probability for Fock
states in the light hadron to have an extra heavy quark pair of mass
$M_Q$ scales as $1/M_Q^2$, which is a result that can be obtained
using the operator product expansion. The result is shown in
Fig.\ref{Inc-NP-IB-LHC}.
Notice that the cross section for inclusive Higgs production from
intrinsic bottom is much higher than the one coming from intrinsic
charm. Although it is true that the Higgs-quark coupling,
proportional to $m_Q$, cancels in the cross section with
$P_{IQ}\propto 1/m_Q^2$, the matrix element between IQ and Higgs
wave functions has an additional $m_Q$ factor. This is because the
Higgs wave function is very narrow and the overlap of the two wave
functions results in $\Psi_{\bar QQ}(0) \propto m_Q$. Thus, the
cross section rises as $m_Q^2$, as we see in the results.

We obtain essentially the same curves for Tevatron energies
($\sqrt{s}=2\TeV$) , although the rates are reduced by a factor of
approximately 3.

Comparing our prediction for inclusive Higgs production coming from
IB with the previous estimate of the g-g fusion cross section, we
see that this high-$x_F$ region is the ideal place to look for Higgs
production coming from intrinsic heavy quarks.

\begin{figure}[tbh]
\includegraphics{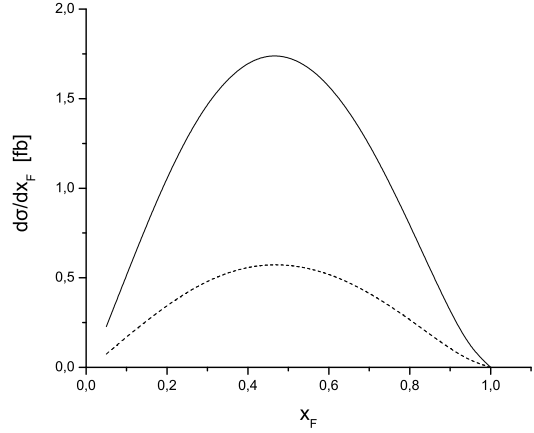}
\begin{center}
\vspace{6.5cm}
\parbox{13cm}
{\caption[Delta] {The cross section of inclusive Higgs production in
$fb$ coming from the perturbative intrinsic charm distribution, at
both LHC ($\sqrt{s}=14\TeV$, solid curve) and Tevatron
($\sqrt{s}=2\TeV$, dashed curve) energies.} \label{Inc-P-IC-LHC}}
\end{center}
\end{figure}

We also show in Fig. \ref{Inc-P-IC-LHC} the results for Higgs
production coming from the perturbative charm distribution. The
magnitude of the production cross section is reduced with respect to
the non-perturbative case.

\begin{figure}[tbh]
\includegraphics{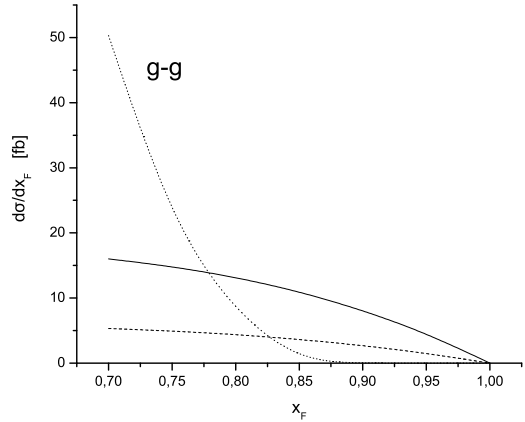}
\begin{center}
\vspace{6.5cm}
\parbox{13cm}
{\caption[Delta] {The cross section of inclusive Higgs production in
$fb$ coming from the perturbative intrinsic bottom distribution, at
both LHC ($\sqrt{s}=14\TeV$, solid curve) and Tevatron
($\sqrt{s}=2\TeV$, dashed curve) energies. For comparison we show
also an estimate of the cross section for gluon-gluon fusion (dotted
curve).} \label{Inc-P-IB-LHC}}
\end{center}
\end{figure}

Nevertheless, Higgs production from perturbative bottom is
substantial for $x_F \ge 0.7$. This is shown in Fig.
\ref{Inc-P-IC-LHC}. Notice that the maximum is shifted towards $x_F
\sim 1/2$ due to the evolution factor. This is just energy loss for
vacuum gluon radiation, which shifts the z-distribution to smaller
z. When we give to a quark a kick as strong as the Higgs mass, it
radiates profusely.

It is also interesting that the very rare fluctuations $|uud t \bar
t>$ in the proton wavefunction from intrinsic top pairs give about
the same contribution as intrinsic bottom. One can see this
explicitly by comparing the two lines of Eq. \ref{10} for intrinsic
bottom and intrinsic top. In spite of the factor $m_Q^2,$ the
$\delta$-dependent function cancels the mass dependence. Indeed, if
$\delta$ is small,  one can expand the bottom line and the first
nonvanishing term is $\delta^2,$ which cancels the $\delta^2$ in the
denominator of Eq. \ref{10}

The contributions of intrinsic bottom and top of perturbative origin
to inclusive Higgs hadroproduction substantially exceed the IC
contribution. Although the probability to find intrinsic heavy
quarks in the proton is very small (the OPE predicts $1/m^2_Q$
scaling for perturbative IB and IT~\cite{Franz:2000ee},) the
stronger Higgs coupling to heavy quarks and the larger projection of
the intrinsic heavy quark Fock state to the Higgs $\bar QQ$ wave
function overcompensate this smallness. As a result, the IB
contribution is about one order of magnitude larger than IC
according to Eq. (1), and the IT contribution turns out to be close
in magnitude to that of IB, but peaked at higher $x_F.$ In fact, the
perturbative IT contribution gives a sharp peak at very high $x_F
\simeq 356/357.$  See also the discussion of
ref.\cite{Brodsky:2006wb} for the case of doubly diffractive Higgs
production.

\section{Summary and Outlook}

The existence of heavy intrinsic Fock states in hadron wavefunctions
is a consequence of the quantum fluctuations inherent to QCD.
The power law fall-off of the probability of intrinsic heavy quark
fluctuations (IQ) is rigorously known in QCD from the operator
product expansion.
In contrast to Abelian theory,  the probability for heavy quark Fock
states such as $|uud Q \bar Q>$ in the proton is only suppressed by
$\Lambda^2/M^2_Q$ because of the contribution of the non-Abelian
operator $G^3_{\mu \nu}$ to the hadronic self-energy. As we have
shown in this paper, the coupling of high-$x$ intrinsic charm and
bottom quark-antiquark pairs in the proton to the Higgs particle
leads to a remarkably large cross section for the hadroproduction of
the Higgs at longitudinal momentum fractions as large as $x_F \simeq
0.8.$ The intrinsic heavy quark mechanism is thus extraordinarily
efficient in converting projectile momentum and energy to Higgs
momentum.

It should be emphasized that the transition of the color-octet $(Q
\bar Q)_{8C}$ intrinsic component of the proton into the Higgs is
due to the coherent interactions of both heavy quarks with the
gluonic field of the other hadron. Since both the $Q$ and $\bar Q$
interact coherently and combine into the color-singlet Higgs, there
is a strong cancellation of the amplitudes. This effect, called
``color transparency", leads to a specific dependence on the Higgs
mass in our calculations.  The light-cone color-dipole description
which we rely upon is well tested in small-$x$ deep-inelastic
scattering and electroproduction of vector mesons, as in
Ref.~\cite{BFGMS,hikt}. Our calculations are done in close analogy
to these reactions.

We predict that the cross section ${d \sigma/d x_F }(p \bar p \to H
X)$ for the inclusive production of the Standard Model Higgs coming
from nonperturbative intrinsic bottom Fock states is of order 50 fb
at LHC energies, peaking in the region of $x_F \sim 0.9$.  Our
estimates indicate that the corresponding cross section coming from
gluon-gluon fusion in this same region is negligible, of order of
0.07 fb, and therefore this peak should be clearly visible. The
coupling of intrinsic bottom to the Higgs is favored over intrinsic
charm  because of its enhanced wavefunction overlap.

The conventional leading-twist contribution to $ p p \to b \bar b X$
arising from gluon fusion $ g g \to b \bar b$ is strongly suppressed
at high $x_F$ because the input gluon distributions are very soft.
Thus the largest background to the Higgs signal arising from IQ $(Q
\bar Q) g \to H \to b \bar b $ is most likely due to the excitation
of the intrinsic $ b \bar b$ Fock states.

In the intrinsic heavy quark subprocess, $ (Q \bar Q)  + g \to H$,
the momentum of both the intrinsic heavy quark and antiquark in the
$|uud Q \bar Q>$ Fock state of the projectile combine to produce the
Higgs at large $x_F.$ This feature of the model can be explicitly
tested and normalized by measuring the production of P-wave heavy
quark Fock states at high $x_F, $ such as $p \bar  p \to \chi_C X.$

We have not included possible extra contributions from intrinsic $t
\bar t$ pairs. The rate for doubly diffractive Higgs production $ p
\bar p \to p + H + \bar p$  at high $x_F$  has been computed in
ref.~\cite{Brodsky:2006wb}.

The results of our paper suggest new strategies for the detection of
the Higgs. The Higgs is produced from intrinsic heavy quarks at high
longitudinal momentum, but its transverse momentum $p^{H}_ \perp$ is
of the order of the heavy quark mass  which is relatively small, and
the sum of the transverse momenta of its decay products will tend to
vanish. Thus if the Higgs decays to two jets such as $b \bar b $,
each jet will be produced with a transverse momentum $\le M_H/2$ and
longitudinal momentum $\simeq P_H/2$. The typical production angle
is then $\theta_{cm} \simeq 2 M_H/  x_F \sqrt s.$ For example,
$\theta_{cm} \simeq  8^o$ for $M_H = 110$ GeV,  $x_F = 0.8$, and
$\sqrt s  = 2$ TeV.  The detection of the Higgs at $x^H_F \sim 0.8$
at Tevatron or the LHC  will thus require forward detection
capabilities.   One can also increase the production angle of  each
jet by lowering the CM energy of the collider or by arranging
collisions with asymmetric beams. Conceivably, the Higgs could even
be produced in $p p$ collisions at RHIC energies.

We also expect a background to Higgs  production at large $x_F$ in
the $H \to b \bar b$ channel arising from the production of
intrinsic $b{ \bar b}$ pairs in QCD.   Since the mean transverse
momenta of quarks inside the intrinsic bottom wavefunction of the
proton is small, the  $b{ \bar b}$ background events which have
large invariant mass come from unequal sharing of quark longitudinal
momentum.   In contrast, the $b{\bar b}$ pairs from the Higgs signal
have a spherically symmetric phase space. One can thus enhance the
signal to background ratio by selecting $b{\bar b}$ jets carrying
large longitudinal momenta. The net signal-to-background ratio in
the $b{\bar b}$ channel is  expected to be similar for the gg fusion
and IQ mechanisms. Furthermore, notice that the background from
$gg\to b\bar b$ consists of pairs where the $b$ and $\bar b$ are
produced at small rapidities and with different invariant mass. This
is not a background for our Higgs signal. The background is thus
reduced by many orders of magnitude. Further kinematic constraints
could provide additional background suppression.

A detailed study of backgrounds will be presented in a separate
publication. The Higgs rapidity distribution for the standard gluon
fusion and bottom annihilation processes has been studied to high
order in ref.~\cite{Ravindran:2007sv}.


There are other interesting tests of intrinsic heavy quark Fock
states, which could be performed at the Tevatron.  For example,
consider $ p \bar p \to W^+ D X$ in which the weak boson $W^+$ is
produced at large $x^W_F \sim 0.4$ in the proton fragmentation
region. The $W^+$ can be created from the pair annihilation subprocesses
$ c \bar d \to W^+$ (which is Cabbibo suppressed), or $c \bar s \to
W^+$. The annihilating $c$ can come from the IC state of the
projectile proton and the $\bar s$ can arise from the strange sea of
the antiproton. The unique signal for this IC process is the
associated production of a forward leading $D^+(\bar c u)$ or
$D^0(\bar c d)$, which is also produced at high $x^D_F$ in the
proton fragmentation region. It is created from the coalescence of
the high-$x$ spectator intrinsic $\bar c$ with a valence quark.
One can similarly test for intrinsic bottom in the reaction $ p \bar
p \to W^- B X$, in which the $W^-$ is produced at large $x^W_F \sim
0.4$ in the proton fragmentation region from the annihilation
subprocess $ b \bar u \to W^-$ (which is CKM-suppressed) or $ b \bar
c \to W^-.$ The annihilating $b$ can come from the IB state of the
projectile proton, and the $\bar c$ can be an anti-charm quark
produced at low-$x$ by gluon splitting. The signal for this IB
process is the associated production of a leading $B^+(\bar b u)$ or
$B^0(\bar b d)$, which is also produced at high $x_B$ in the proton
fragmentation region. It is created from the coalescence of the
spectator intrinsic $\bar b$ with a valence quark. Other tests
include the reactions $ g c  \to Z^0/\gamma +  c $ where the $Z^0$
plus the tagged final-state charm jet have large
$x_F$~\cite{Pumplin:2007wg}. It has also been found that intrinsic
bottom could even contribute significantly to exotic processes such
as neutrino-less $\mu^- - e^-$ conversion in nuclei~\cite{KSS}. All
of the above mentioned tests require detailed calculations which
consider other possible production mechanisms and backgrounds.

If the Higgs boson were detected at large Feynman $x$, it would not
only confirm the structure of the intrinsic heavy-quark Fock states
as predicted by QCD, but also the fundamental coupling strengths of
the Higgs bosons to heavy quarks.

\bigskip

{\bf Acknowledgments:}  This work was supported in part by the
Department of Energy under contract number DE-AC02-76SF00515, by
Fondecyt (Chile) grant 1050589, and by DFG (Germany)  grant
PI182/3-1. We thank  Paul Grannis, Jacques Soffer, and Alfonso
Zerwekh for helpful conversations.

\end{document}